\documentclass[runningheads]{cl2emult}

\usepackage{makeidx}  % allows index generation
\usepackage{graphicx} % standard LaTeX graphics tool
\usepackage{subeqnar} % subnumbers individual equations
\usepackage{multicol} % used for the two-column index
\usepackage{cropmark} % cropmarks for pages without
\usepackage{math}     % placeholder for figures
\makeindex            % used for the subject index
\begin{document}
\title*{\sffamily Physics of Personal Income}
%\title*{Focusing of a Parallel Beam to Form\protect\newline a Point
%in the Particle Deflection Plane}
%
%
\toctitle{Focusing of a Parallel Beam to Form a Point
\protect\newline in the Particle Deflection Plane}
% allows explicit linebreak for the table of content
%
%
%%%%%%% ¤³¤³¤Ï¾Êά?
%\titlerunning{Physics of Personal Income}
%\titlerunning{Focusing of a Parallel Beam}
% allows abbreviation of title, if the full title is too long
% to fit in the running head
%
\author{Wataru Souma}
%\and Roger Temam\inst{2}
%\and Jeffrey Dean\inst{2}
%\and David Grove\inst{1}
%\and Craig Chambers\inst{2}
%\and Kim~B.~Bruce\inst{2}
%\and Elsa Bertino\inst{1}}
%
%%%%%%% ¤³¤³¤Ï¾Êά?
%\authorrunning{Wataru Souma}
%\authorrunning{Ivar Ekeland et al.}
% if there are more than two authors,
% please abbreviate author list for running head
%
%
%\institute{Graduate School of Human and Environmental Studies,
%Kyoto University, Kyoto 606-8501, Japan
%}
\institute{ATR Human Information Science Laboratories,
Kyoto 619-0288, Japan
%\footnote{The main part of this article has been performed at
%Kyoto Univ}
%the Faculty of Integrated Human Studies,
%Kyoto University, Kyoto 606-8501, Japan.}
}
%\institute{Princeton University, Princeton NJ 08544, USA
%\and Universit\'{e} de Paris-Sud,
%     Laboratoire d'Analyse Num\'{e}rique,
%     B\^{a}timent 425,\\
%     F-91405 Orsay Cedex, France}

\maketitle              % typesets the title of the contribution

\vspace{12pt}
%\begin{abstract}
\noindent{\bfseries Summary.} We report empirical studies on the personal
income distribution,
and clarify that the distribution pattern of the lognormal
with power law tail is the universal structure.
We analyze the temporal change of Pareto index and Gibrat index
to investigate the change of the inequality of
the income distribution.
In addition some mathematical models which are proposed to
explain the power law distribution
are reviewed.
%\end{abstract}

\vspace{12pt}

\vspace{24pt}
\noindent{\large\sffamily\bfseries 1. Introduction}
%\section{Introduction}
\vspace{12pt}

\noindent
A study of the personal income distribution
has important meaning in the econophysics,
because the personal income is a basic ingredient
of the economics.

The study of the personal income has long history
and many investigations have been done.
The starting point is about one hundred years ago when
V.~Pareto proposed the power law distribution of the personal income
(Pareto 1897).
He analyzed the distribution of the personal income
for some countries and years, and found that
the probability density function $p(x)$ of the personal income $x$
is given by
\[
p(x)=Ax^{-(1+\alpha)},
\]
where $A$ is the normalization constant.
This power law behavior is called Pareto law and
the exponent $\alpha$ is named Pareto index.
This is a classic example of fractal distributions,
and observed in many self-organizing systems.
If Pareto index has small values, the personal income is unevenly distributed.
Some examples of Pareto index are summarized in Table.~1 (Badger 1980).

However, it is well known
that Pareto law is only applicable to
the high income range.
It was clarified by R.~Gibrat that the distribution takes the form of the
lognormal in the middle income range (Gibrat 1931).
As is well known, in this case, the probability density
function is given by
\[
p(x)=\frac{1}{x\sqrt{2\pi\sigma^2}}\exp
\left[-\frac{\log^2\left(x/x_0\right)}{2\sigma^2}\right],
\]
where $x_0$ is a mean value and $\sigma^2$ is a variance.
Sometimes $\beta\equiv1/\sqrt{2\sigma^2}$ is called Gibrat index.
Since the large variance means the global distribution of the income,
the small $\beta$ corresponds to the uneven distribution of the
personal income.

\begin{table}[t]
\caption{Examples of Pareto index $\alpha$ for some countries and years
(Badger 1980, with permission from Taylor \& Francis Ltd.).}
\renewcommand{\arraystretch}{1}
\setlength\tabcolsep{15pt}
\begin{tabular}{@{}lrclc}
\hline\noalign{\smallskip}
Country & &$\alpha$ & Country & $\alpha$\\
\noalign{\smallskip}
\hline
\noalign{\smallskip}
England&(1843)   &1.50& Perugia(city)            & 1.69\\
       &(1879-80)&1.35& Perugia(country)         & 1.37\\
       &(1893-94)&1.50& Ancona,Arezzo,& 1.32\\
Prussia&(1852)   &1.89& Parma,Pisa & \\
       &(1876)   &1.72& Italian cities           & 1.45\\
       &(1881)   &1.73& Basel                    & 1.24\\
       &(1886)   &1.68& Paris(rents)             & 1.57\\
       &(1890)   &1.60& Florence                 & 1.41\\
       &(1894)   &1.60& Peru(at the end of       &1.79 \\
Saxony &(1880)   &1.58& 18th century)            & \\
       &(1886)   &1.51&                          & \\
Augsburg&(1471)  &1.43&                          & \\
        &(1498)  &1.47&                          & \\
        &(1512)  &1.26&                          & \\
        &(1526)  &1.13&                          & \\
\noalign{\smallskip}
\hline
\noalign{\smallskip}
\end{tabular}\\
\label{Tab1b}
\end{table}

\begin{figure}[t]
\centering
\includegraphics[width=.67\textwidth]{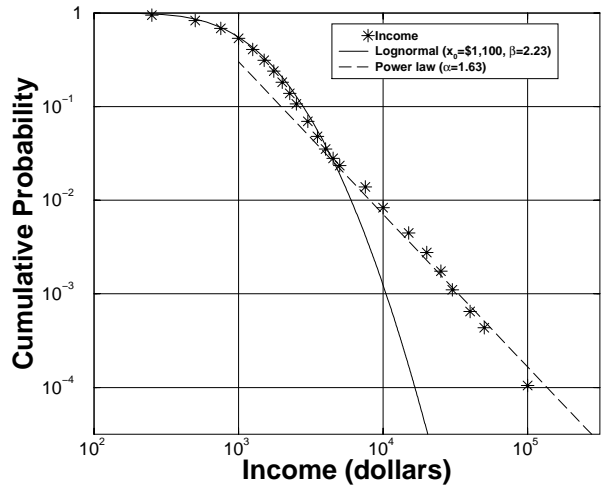}
\caption[]{The power law and lognormal fits to the 1935-36 U.S. income
data. The solid line represents the lognormal fit with $x_0=\$1,100$
and $\beta=2.23$. The straight dashed line represents the
power law fit with $\alpha=1.63$
(Badger 1980, with permission from Taylor \& Francis Ltd.).}
\label{fig1}
\end{figure}

The lognormal distribution with power law tail
for the personal income is rediscovered by 
Badger (1980) and Montroll and Shlesinger (1980)
\footnote{Recently the exponential
distribution of the personal
income has reported by Dr\u{a}gulescu and Yakovenko (2000).}.
Those investigations were performed for
the 1935-36 U.S. income data, and
confirmed that
the top 1\% of the distribution follows Pareto law with $\alpha=1.63$,
and the other follows the lognormal distribution with
$x_0=\$1,100$ and $\beta=2.23$.
The distribution is shown in Fig.~\ref{fig1}.
In this figure, we take the horizontal axis as the logarithm of the income
with the unit of dollars,
and the vertical axis as the logarithm of the cumulative probability $P(x\leq)$.
The cumulative probability is the probability finding the person
with the income greater than or equals to $x$, and
defined by $P(x\leq)\equiv\int_{x}^\infty dy p(y)$
in the continuous notation.
In other words the cumulative probability is the rank normalized by
the total number of individuals.
In this figure the dashed line and the thin solid line are
the fitting of the power law and the lognormal functions respectively.

From the age of J.J.~Rousseau,
one of the subject
of the social science is the theory of the inequality.
Many indexes specifying the
unevenness of the income distribution have been proposed
in the economics.
Among them,
Gini index is well known and frequently used.
Although Gini index is a useful measure of the uneven distribution,
this index has no attraction from the physical point of view.
This is because the manipulation deriving Gini index hides the
mechanism explaining the distribution of the personal income.

Though many investigations of the personal income distribution
have been performed, data sets are all old.
Hence to reanalyze the income distribution by the recently
high quality data is meaningful.
In a previous article (Aoyama et al. 2000 and see the chapter
by H. Aoyama, this volume), we analyzed the
personal income distribution of Japan in the year 1998,
and clarified that the high income range 
follows Pareto law with $\alpha=2.06$.
However, this analysis is incomplete
for the middle income region, because
the data set for this region is sparse.
Hence, the firstly we
gain the overall profile of the personal income distribution
of Japan in that year.
In addition we perform same analysis for the Japanese personal
income in another years, and compare them with the result of
Badger (1980) and Montroll and Shlesinger (1980).
From these studies we deduce the universal structure of the
personal income distribution. 

Secondly we focus on the temporal change of $\alpha$ and
$\beta$
\footnote{The temporal change of $x_0$ and the correlation between
$x_0$ and the Gross Domestic Product (GDP) are summarized by Souma (2000).}.
Although these indexes have been estimated in many countries and many years
as shown in Table.~1,
the succeeding change of these indexes is
not well known. Hence the investigation of the temporal change
of these indexes has important meaning.

Lastly we review models based on the stochastic process
explaining the power law distribution.
Though these models have not been developed to explain the personal
income distribution, the useful information is contained.

\vspace{24pt}
\noindent{\sffamily\bfseries\large
2. Universal structure of the personal income distribution}
%2. Lognormal distribution with power law tail}
%\section{Lognormal distribution with power law tail}
\vspace{12pt}

\noindent
To obtain the overall profile of the personal income distribution
of Japan in the year 1998, we use three data sets;
{\itshape income-tax data}, {\itshape income data} and
{\itshape employment income data}.

The {\itshape income-tax data} is only
available for the year 1998.
This is a list of the 84,515 individuals who paid the income-tax of
ten million yen or more in that year.

The {\itshape income data} contains the person who
filed tax return individually, and
a coarsely tabulated data.
We analyze this data over the 112 years 1887-1998 in this article.
The data is publicly available from the Japanese Tax Administration (JTA)
report,
and the recent record is on the web pages of the JTA. 

The {\itshape employment income data} is the 
sample survey for the salary persons working in the private enterprises,
and does not contain the public servants and the persons with
daily wages.
This data is coarsely tabulated as same as the {\itshape income data}.
We analyze this data over the 44 years 1955-98 in this article.
This is publicly available from the JTA report and
recent record is available on the same web pages for the
{\itshape income data}.
The distribution is recorded with the unit
of thousand people from the year 1964.

To gain the overall profile of the personal income distribution,
we connect these data sets with following rules.
\begin{enumerate}
\item We use adjusted {\itshape income-tax data} in the range $50\le x$,
where $x$ has the unit of million yen.
We translate the income-tax $t$ to the income $x$ as $t=0.3 x$
in this range
(Aoyama et al. 2000 and see the chapter by H. Aoyama, this volume).
\item We only use the {\itshape income data} in the range $20\le x<50$. 
This is because all persons with income
greater than 20 million yen must file tax return individually
under the Japanese tax system from the year 1965.
Hence individuals with employment income greater
than 20 million yen must file tax return
individually, and are counted in the {\itshape income data}.
\item We sum up {\itshape income data} and
{\itshape employment income data} in the range $x<20$.
Although the double counted persons exist,
detailed information to remove this ambiguity is impossible.
\end{enumerate}
From these process we have the data for 51,493,254 individuals,
about $80\%$ of all workers in Japan.

\begin{figure}[t]
\centering
\includegraphics[width=.67\textwidth]{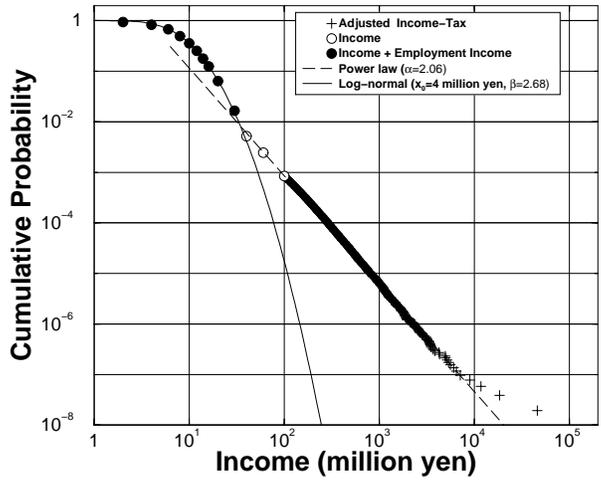}
\caption[]{The power law and lognormal fits to the 1998 Japanese
income data.
The tin solid line represents the lognormal fit with $x_0=4$ million yen
and $\beta=2.68$. The straight dashed line represents the
power law fit with $\alpha=2.06$.}
\label{fig2}
\end{figure}

The distribution for the year 1998 is shown in Fig.~\ref{fig2}.
In this figure, we take the horizontal axis as the logarithm of the income
with the unit of million yen,
and the vertical axis as the logarithm of the cumulative probability.
The bold solid line corresponds to the adjusted {\itshape income-tax data}.
Open circles emerge from only the {\itshape income data}, and
filled circles derived from the sum of the {\itshape income data} and
the {\itshape employment income data}.
The dashed line and the thin solid line are the fitting of the power law
and the lognormal respectively.
We recognize from this figure that the top $1\%$ of the distribution
follows Pareto law with $\alpha=2.06$.
On the other hand $99\%$ of the distribution follows the
lognormal distribution with
$x_0=4$-million yen and $\beta=2.68$.
The change from the lognormal to the power law
does not occur smoothly, and this is also observed in Fig.~\ref{fig1}.
As will be shown later, this discontinuous change is observed
for another years.

Although the detailed data for the high income
(i.e, {\itshape income-tax data})
is only available for the
year 1998, the overall profile of the
distribution can be gained from the {\itshape income data}
and the {\itshape employment income data}
as recognized from Fig.~\ref{fig2}.
Moreover the value of $\alpha$ is
available from only the {\itshape income data},
open circles in Fig.~\ref{fig2}.
Hence the {\itshape income data} should give an idea of
the value of $\alpha$.
We use previous rules to gain the overall profile of the personal
income distribution for these years.

\begin{figure}[t]
\centering
\includegraphics[width=.49\textwidth]{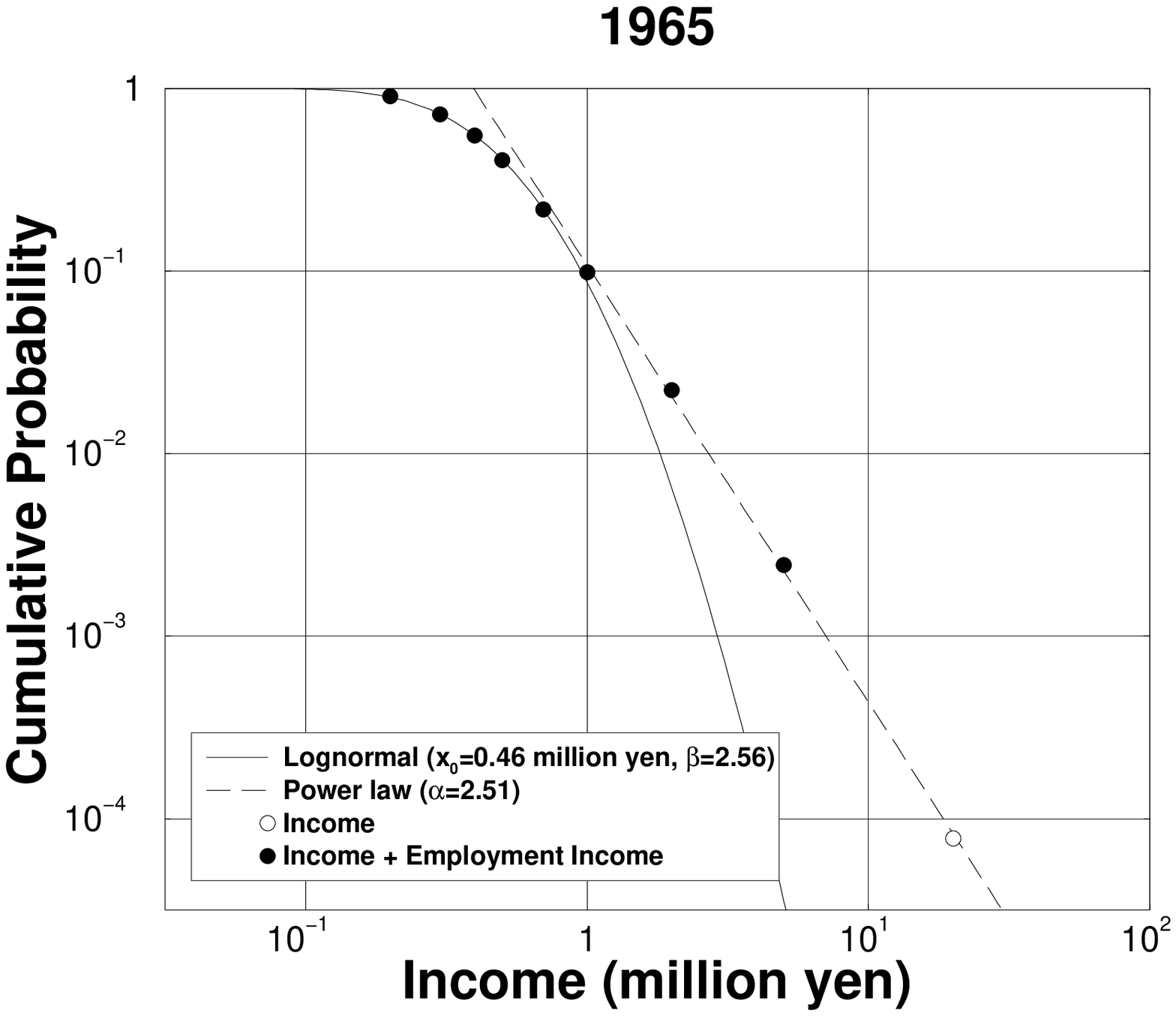}
\includegraphics[width=.49\textwidth]{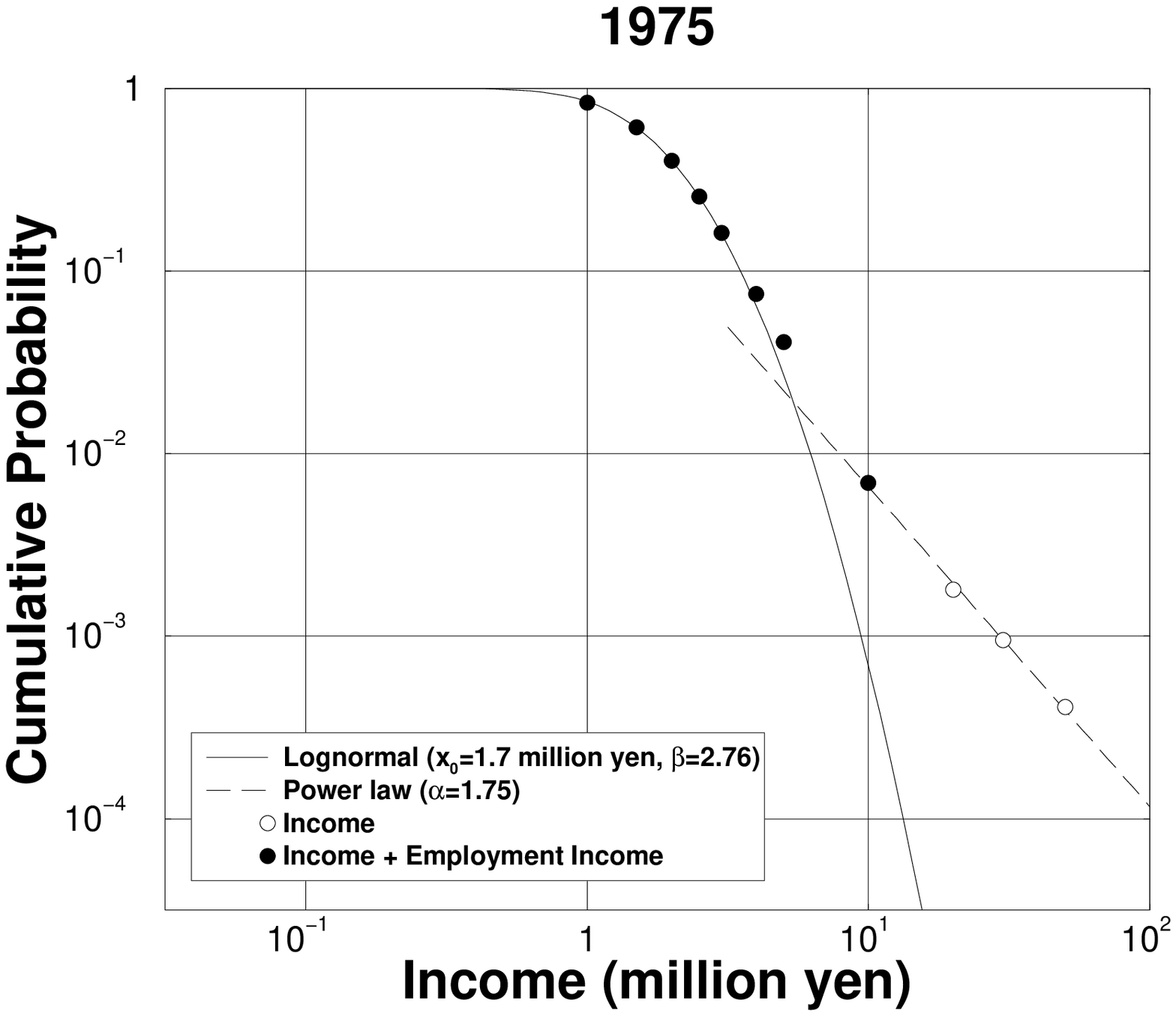}
\includegraphics[width=.49\textwidth]{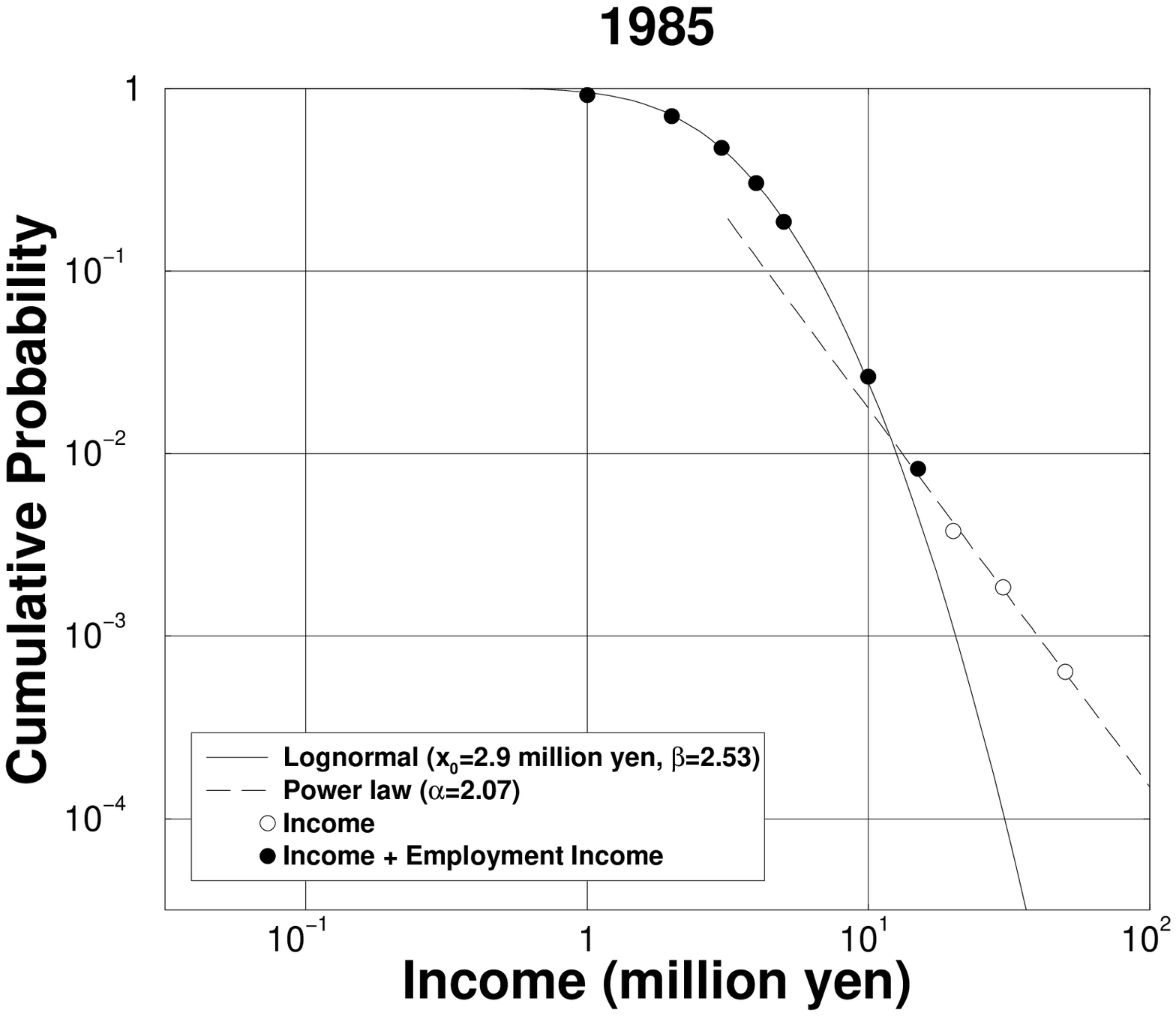}
\includegraphics[width=.49\textwidth]{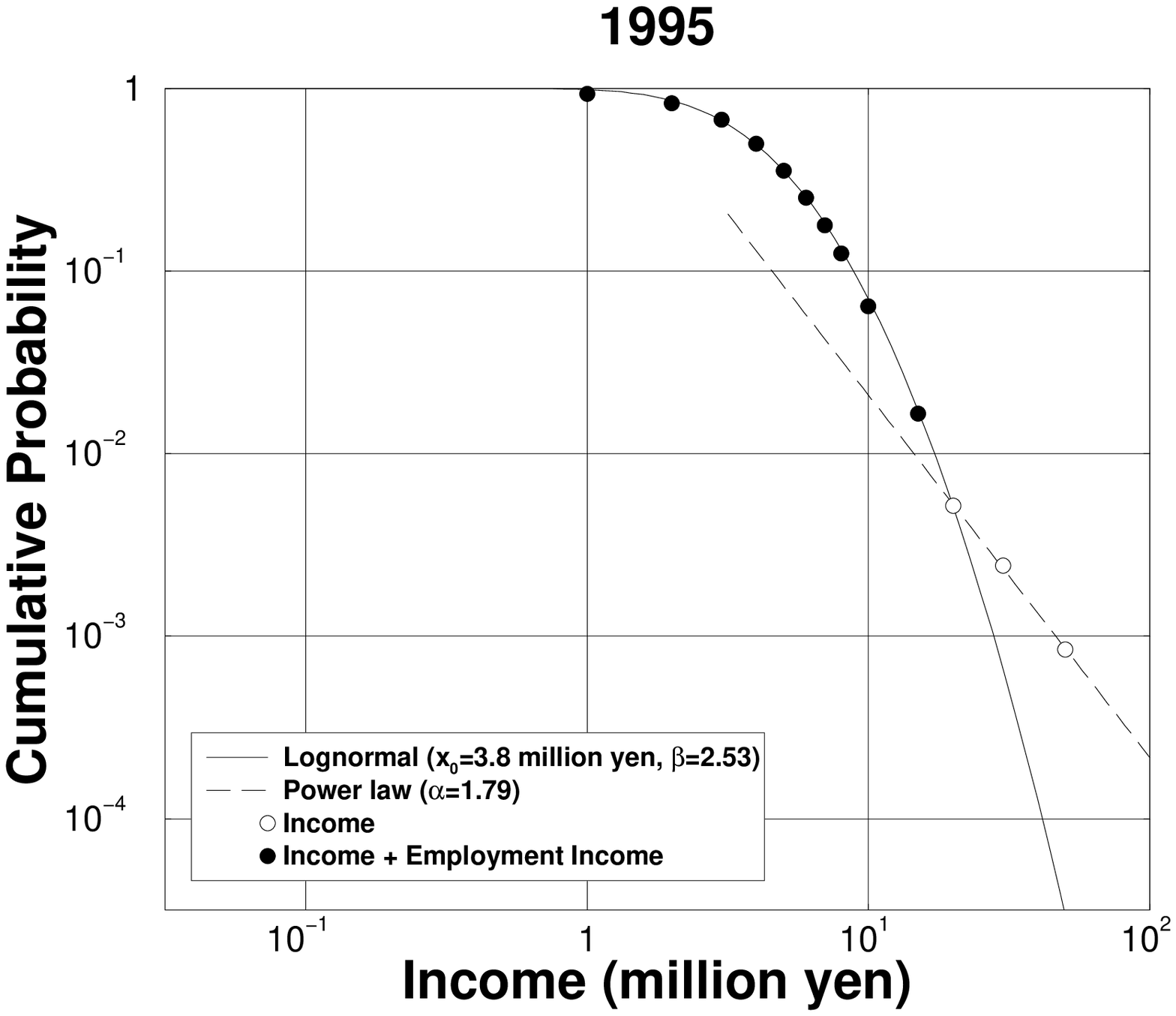}
\caption[]{
The power law and lognormal fit to the 1965, 1975, 1985 and 1995
Japanese income data.
The solid line represents the lognormal fit and the straight dashed
line represents the power law fit.
}
\label{fig3}
\end{figure}

The distributions for the years 1965, 1975, 1985 and
1995 are shown in Fig.~\ref{fig3}.
Solid lines in Fig.~\ref{fig3} are the lognormal fit for the
middle income range and
dashed lines are the power law
fit for the high income range.
We recognize that less than top 10\% of the distribution
is well fitted by Pareto law and greater than 90\% of
the distribution follows the lognormal distribution.
However the slope of each dashed lines and
the curvature of each solid lines differ from each other.
Hence Pareto index and Gibrat index differ from year to year.
The movement of the distribution toward the right direction consists
in the increase of the mean income, and
is characterized by the change of $x_0$.
If we normalize the income by the
inflation or deflation rate, this movement may be deamplified.
However, even if those manipulations are applied, the profile of the
distribution is not modified.
As stated before the discontinuous change from lognormal to
power law are observed for the year 1975, 1985 and 1995 in Fig.~\ref{fig3}.
However the reason of this is not known.

Although some ambiguities and unsolved problems exist,
we can confirm
that the distribution pattern of the personal income is expressed
as the lognormal with power law tail.
This distribution pattern
coincides with the result of
Badger (1980) and Montroll and Shlesinger (1980).
Hence we can say that the lognormal with power law tail
of the personal income distribution is a universal structure.
However
the indexes specifying the distribution
differ from year to year as recognized from
Figs.~\ref{fig2} and~\ref{fig3}. 
We therefore study the temporal change of these indexes in the next section.

\vspace{24pt}
\noindent{\sffamily\bfseries\large
3. The temporal change of the distribution}
\vspace{12pt}
%\section{The temporal change of the distribution}

\begin{figure}[t]
\centering
\includegraphics[width=.67\textwidth]{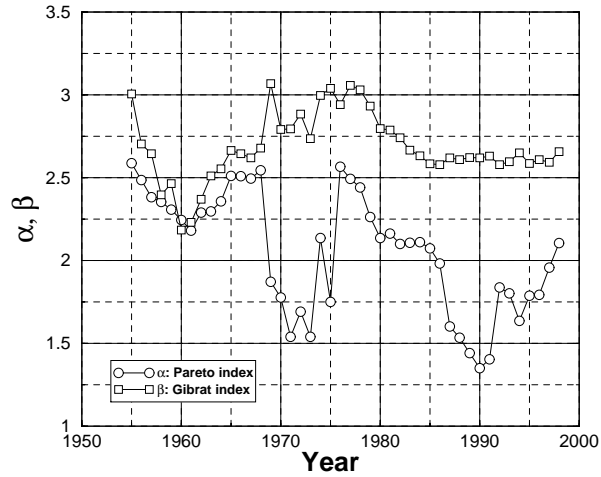}
\caption[]{
The temporal change of $\alpha$ and $\beta$ in Japan over the 44 years
1955-98.
Open circles represent the change of $\alpha$ and open squares
represent that of $\beta$.
}
\label{fig4}
\end{figure}

We consider the change of $\alpha$ and $\beta$  in Japan
over the 44 years 1955-98.
We have Fig.~\ref{fig4} from the numerical fit of the distribution.
In this figure the horizontal
axis is the year and the vertical axis
is the value of $\alpha$ and $\beta$.
Open circles and squares correspond to $\alpha$ and
$\beta$ respectively.
It is recognized from this figure that these
indexes correlate with each other around the year 1960 and 1980.
However these quantities
have no correlation in the beginning of the 1970s and after the year 1985.
In the range where $\alpha$ and $\beta$ change independently,
the strongly changing index is $\alpha$.
Especially $\beta$ stays almost same value after the year 1985.
This means that the variance of the middle income is not changing.
From these behaviors of $\alpha$ and $\beta$, we can consider
that there are some factors causing
no correlation between
$\alpha$ and $\beta$, and mainly effecting to $\alpha$
\footnote{Correlations between
$\alpha$ and the land price index and the TOPIX
are summarized by Souma (2000).}.

\begin{figure}[t]
\centering
\includegraphics[width=.67\textwidth]{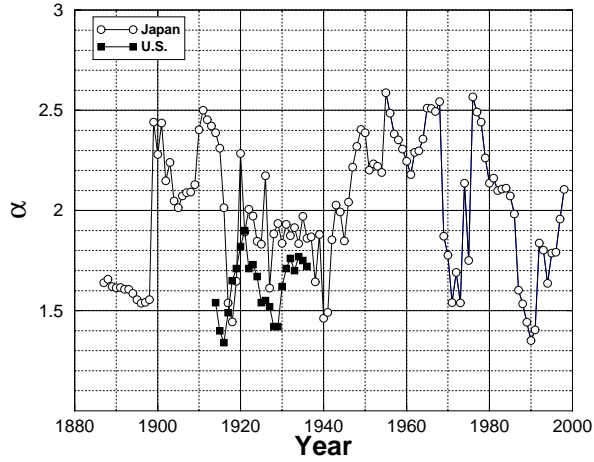}
\caption[]{The temporal change of $\alpha$.
Open circles represent the change of $\alpha$ in Japan over
the 112 years 1887-1998.
Filled squares represent the change of $\alpha$ in U.S. over
the 23 years 1914-36 (Badger 1980,
with permission from Taylor \& Francis Ltd.).
}
\label{fig5}
\end{figure}

As mentioned previously, $\alpha$
is mainly derived from the {\itshape income data}.
Hence the idea of the change of $\alpha$ can be
obtained over the 112 years 1887-1998 for the Japanese income
distribution.
The data analysis derives open circles in Fig.~\ref{fig5}.
In this figure the horizontal axis is the year and the
vertical axis is the value of $\alpha$.
The mean value of Pareto index is $\bar\alpha=2$, and $\alpha$
fluctuates around it.
This
is worth to compare with the case of the Japanese
company size and that of the debts of bankrupt companies
(Okuyama et al. 1999, Aoyama et al. 2000, and see the chapter by
M. Katori and T. Mizuno, this volume).
In these cases the distribution follows the power law with $\alpha=1$;
Zipf's law. 
Filled squares represent the change of $\alpha$ in U.S. over
the 23 years 1914-36 (Badger 1980).
The interesting observation is that the behaviors of $\alpha$ in Japan
and that in U.S. almost coincide.

\vspace{24pt}
\noindent{\sffamily\bfseries\large
4. Mathematical models}
\vspace{12pt}
%\section{Mathematical models}

The most simple model considered to explain the income
distribution is the pure multiplicative stochastic process (MSP).
This model is defined by
\[
x(t+1)=a(t)x(t),
\]
where
$a(t)$ is the positive random variables.
Hence if this process is iteratively applied, we have
\[
x(t+1)=a(t)\cdot a(t-1)\cdots x(0).
\]
The logarithm of this equation derives
\[
\log x(t+1)=\log a(t) +\log a(t-1) +\cdots+\log x(0). 
\]
Thus $\log x(t+1)$ follows the normal distribution,
and $x(t+1)$ does the lognormal one.
Though this pure MSP well explain the lognormal distribution,
this derives the monotonically increasing variance contrary
to the empirical observation shown in Fig.~\ref{fig4},
and fails to explain the power law tail.

Some models have been proposed to beyond the pure MSP.
It has been shown that boundary constraints (Levy and Solomon 1996) and
additive noise (Kesten 1973, Sornette and Cont 1997,
Takayasu et al. 1997, Sornette 1998)
are able to induce the MSP to generate power law.
The MSP with boundary constraints
is defined by the same equation of the pure MSP.
The difference from them consists in constraints:
\[
\langle\log a(t)\rangle<0,\;\;\;0<x_{\textrm{\scriptsize m}}<x(t),
\]
where $x_{\textrm{\scriptsize m}}$ is the poverty bound.
These constraints express that the net drift to
$x(t\rightarrow\infty)\rightarrow-\infty$ is balanced by the reflection
on the reflecting barrier located at $0<x_{\textrm{\scriptsize m}}$.
In this case Pareto index is given by
$\alpha=1/(1-x_{\textrm{\scriptsize m}})$.

The MSP with additive noise is defined by
\[
x(t+1)=a(t)x(t)+b(t),
\]
where $a(t)$ and $b(t)$ are positive independent random variables,
and with the constraint $\langle\log a(t)\rangle<0$.
In this case Pareto index is given by $\langle a^\alpha \rangle=1$
independently of the distribution of $b(t)$.

The equivalence of these models
has been clarified by Sornette and Cont (1997),
and the generalization of them
has been given by
\[
x(t+1)=e^{f(x(t),\{a(t),b(t),\cdots\})}a(t)x(t),
\]
where $f(x(t),\{a(t),b(t),\cdots\})\rightarrow0$ for
$x(t)\rightarrow\infty$
and $f(x(t),\{a(t),b(t),\cdots\})\rightarrow\infty$ for
$x(t)\rightarrow0$.
The MSP with boundary constraints is the special case
$f(x(t),\{a(t),b(t),\cdots\})=0$ for $x_{\textrm{\scriptsize m}}<x(t)$ and
$f(x(t),\{a(t),b(t),\cdots\})=\log(\frac{x_{\textrm{\tiny m}}}
{a(t)x(t)})$ for
$x(t)\leq x_{\textrm{\scriptsize m}}$.
The MSP with additive noise is the special case
$f(x(t),\{a(t),b(t),\cdots\})=\log(1+\frac{b(t)}{a(t)x(t)})$.

Though these models well explain the emergence of the power law distribution,
they are incomplete when we consider the application of them to
the personal income distribution.
This is because interactions between agents are
not included in these models.
Hence interacting MSPs are developed by
several articles.
One is based on so-called `directed polymer' problem
(Bouchaud and M\'{e}zard 2000)
and the other on the generalized Lotka Voltera model
(Solomon and Levy 1996, Biham et al. 1998).
The former model is proposed to explain the wealth distribution of
individuals and companies, and defined by
\[
\frac{dx_i(t)}{dt}=\eta_i(t)x_i(t)+\sum_{j(\neq i)}J_{ij}(t)x_j(t)
-\sum_{j(\neq i)}J_{ji}(t)x_i(t),
\]
where $\eta_i(t)$ is a Gaussian random variable of mean $m$ and variance
$2\sigma^2$, which describes the spontaneous growth or decrease of
wealth due to investment in stock markets, housing, etc.
The terms involving the (asymmetric) matrix $J_{ij}(t)$ describe the
amount of wealth that agent $j$ spends buying the production of agent
$i$ (and vice versa).
Under the mean field approximation, i.e., $J_{ij}(t)=J/N$, where $N$ is the
total number of agents, the stationery solution
has the power law tail
with $\alpha=1+J/\sigma^2$ in the limit of $N\rightarrow\infty$.
It is also clarified that
the dependence of $\alpha$ on $J/\sigma^2$ is not modified even if
we abandon the mean field approximation.

The latter model is defined by
\[
x_i(t+1)-x_i(t)=\left[\varepsilon_i(t)\sigma_i+c_i(x_1,x_2,\cdots,x_N,t)\right]
x_i(t)+a_i\sum_{j}b_jx_j(t),
\]
where $\varepsilon_i(t)$ are the random variables with 
$\langle\varepsilon_i(t)\rangle=0$ and
$\langle\varepsilon_i^2(t)\rangle=1$.
Hence the term in the bracket of the LHS first term corresponds
to the random variable with mean $c_i(x_1,x_2,\cdots,x_N,t)$
and the standard deviation $\sigma_i$.
Here $c_i$ express the systematic endogenous and exogenous trends in the
returns. The last term in the RHS represents the wealth redistributed,
and $a_i$ and $b_i$ represent the amount of wealth redistributed to the
individuals and the contribution of the individuals to the total 
wealth respectively.
If we take $\sigma_i^2=\sigma^2, a_i=a, b_i=1/N,
c_i(x_1,x_2,\cdots,x_N,t)=c(x_1,x_2,\cdots,x_N,t)$ and $N\rightarrow\infty$,
the stationery solution has the power law tail with
$\alpha=1+2a/\sigma^2$,
which coincides with the result of previous model.
The finite $N$ corrections have also been calculated.

Though these models have not been developed to explain the
personal income distribution, the useful information to construct
the mathematical models explaining the personal income distribution
is contained.

\vspace{24pt}
\noindent{\sffamily\bfseries\large
References}

\vspace{12pt}
\begin{small}
\noindent
Aoyama H, et al (2000)
Pareto's law for income of
individuals and debt of
bankrupt

\vspace{-1pt}
companies. Fractals 8:293--300.

\vspace{-1pt}
\noindent
Badger WW (1980)
An entropy-utility model for the size distribution of income.

\vspace{-1pt}
In:West. BJ (Ed) Mathematical models as a tool for the social
science.
Gordon

\vspace{-1pt}
and Breach, New York, pp 87--120.

\vspace{-1pt}
\noindent
Biham O, et al (1998)
Generic emergence of power law distributions
and L\'{e}vy

\vspace{-1pt}
-stable
intermittent fluctuations in discrete logistic systems.
Phys
Rev E58:

\vspace{-1pt}
1352--1358.

\vspace{-1pt}
\noindent
Bouchaud JP, M\'ezard M (2000)
Wealth condensation in a simple model of
economy.

\vspace{-1pt}
Physica A282:536--545.

\vspace{-1pt}
\noindent
Dr\u{a}gulescu A, Yakovenko VM(2000)
Statistical mechanics of money.
Eur Phys J

\vspace{-1pt}
B17:723--729.

\vspace{-1pt}
\noindent
Gibrat R (1931) Les in\'egalit\'s \'economiques. Paris, Sirey.

\vspace{-1pt}
\noindent
Kesten H (1973) Random difference equations and renewal theory for
products of

\vspace{-1pt}
random matrices.
Acta Math 131:207--248.

\vspace{-1pt}
\noindent
Levy M, Solomon S (1996)
Power laws are logarithmic Boltzman laws.
Int
J Mod

\vspace{-1pt}
Phys C7:595--601.

\vspace{-1pt}
\noindent
Montroll EW, Shlesinger MF (1983)
Maximum entropy formalism, fractals,
scaling

\vspace{-1pt}
phenomena, and $1/f$
noise: a tale of tails.
J Stat Phys 32:209--230.

\vspace{-1pt}
\noindent
Okuyama K, et al (1999)
Zipf's law in income distribution
of companies.
Physica

\vspace{-1pt}
A269:125--131.

\vspace{-1pt}
\noindent
Pareto V (1897)
Cours d'\`{e}conomique politique.
Macmillan, London.

\vspace{-1pt}
\noindent
Sornette D (1998)
Multiplicative processes and power laws.
Phys Rev E57:
4811--

\vspace{-1pt}
4813.

\vspace{-1pt}
\noindent
Souma W (2000) Universal structure of the personal income distribution.
cond-

mat/0011373.

\vspace{-1pt}
\noindent
Solomon S, Levy M (1996)
Spontaneous scaling emergence in generic
stochastic

\vspace{-1pt}
systems.
Int J Mod Phys C7:745--751.

\vspace{-1pt}
\noindent
Sornette D, Cont R(1997)
Convergent multiplicative processes repelled from
zero:

\vspace{-1pt}
power laws and truncated power laws. J Phys I7:431--444.

\vspace{-1pt}
\noindent
Takayasu H, et al (1997)
Stable infinite variance fluctuations in
randomly amplified

\vspace{-1pt}
Langevin systems.
Phys Rev Lett 79:966--969.
\end{small}
%\vfill
%INDEX%%%%%%%%%%%%%%%%%%%%%%%%%%%%%%%%%%%%%%%%%%%%%%%%%%%%%%%%%%%%%%%
\clearpage
\addcontentsline{toc}{section}{Index}
\flushbottom
\printindex
%%%%%%%%%%%%%%%%%%%%%%%%%%%%%%%%%%%%%%%%%%%%%%%%%%%%%%%%%%%%%%%%%%%%%
\end{document}